# Anomalous Lattice Dynamics of Mono-, Bi-, and Tri-layer WTe$_2$

*Younghee Kim, [+] Young In Jhon, [+] June Park, [+] Jae Hun Kim, Seok Lee, and Young Min Jhon\**

Sensor System Research Center, Korea Institute of Science and Technology
Seoul, 136-791, Republic of Korea

[+]These authors contributed equally to this work.

\*E-mail: ymjhon@kist.re.kr

**Tungsten ditelluride (WTe$_2$) is a layered material that exhibits excellent magnetoresistance and thermoelectric behaviors, which are deeply related with its distorted orthorhombic phase that may critically affect the lattice dynamics. Here, for the first time, we present comprehensive characterization of the Raman spectroscopic behavior of WTe$_2$ from bulk to monolayer using experimental and computational methods. We discover that mono and bi-layer WTe$_2$ can be easily identified by Raman spectroscopy since double or single Raman modes that are observed in higher-layer WTe$_2$ are substantially suppressed in the monolayer and bilayer WTe$_2$, respectively. In addition, different from hexagonal metal dichalcogenides, the frequency of in-plane mode $A_1^7$ of WTe$_2$ remains almost constant as the layer number decreases, while the other Raman modes consistently blueshift. First-principles calculation validates the experiments and reveals that the negligible shift of the $A_1^7$ mode is attributed to the lattice vibration along the tungsten chains that make WTe$_2$ structurally one-dimensional.**

## 1. Introduction

Graphene success has shown that it is possible to create stable single- or few layers of van der Waals materials, and also that these two-dimensional materials can exhibit exotic condensed matter physics that are absent in the bulk counterparts.[1,2] Recently, atomic-layered transition-metal dichalcogenides $MX_2$ where M is a transition metal (Mo and W) and X is a chalcogen (S, Se, and Te)[3,4] have gained increasing attention for potential applications in electronics and photonics due to their sizable band gaps that change from indirect to direct in single layers[5-8] and superb optoelectronic properties that are suitable for topological insulator application for femtosecond pulse laser.[9-11] However, critical layer dependence of the electronic properties of these materials has required the development of the method for their layer-controlled preparation as well as a precise determination of the layer number.

As shown in our previous work, optical analysis methods can provide a variety of benefits for characterizing two-dimensional materials.[12] In particular, after successful application of Raman analysis in determining the number of layers in graphene,[13] Raman spectroscopy has widely been used to determine the number of layers in the other types of two-dimensional materials including hexagonal transition metal dichalcogenides (2H-$MX_2$) such as $MoS_2$,[14,15] $MoSe_2$,[16,17] $WS_2$,[18,19] $WSe_2$,[16,19] and $MoTe_2$,[20] which consist of hexagonally arranged metal atoms that are sandwiched between two layers of chalcogen atoms.

Lee *et al.*[14] reported that the frequency of the Raman mode $A_{1g}$ decreases and that of the Raman mode $E_{2g}^1$ increases as the layer number decreases, which can be explained by van der Waals interlayer coupling, Coulombic interactions, and stacking-induced changes of the intralayer bonding. They indicated that the frequency difference between these two modes can offer a convenient and reliable means for determining layer thickness with atomic-layer precision.The Raman studies of 2H-$MX_2$ have been frequently accompanied by first-principles calculations for a deep understanding of their lattice dynamics. Lianbo *et al.* studied the Raman spectroscopies of few-layer MoS$_2$ and WS$_2$ using density functional theory (DFT) calculations.[21] They showed that vertically-stacked MoS$_2$/WS$_2$ heterostructures (up to four layers) can exhibit distinctly different Raman spectra with respect to the frequency and intensity depending on the stacking order, which can be explained by the changes in dielectric screening and interlayer interaction. They also supposed that non-conclusive experimental observation for the Raman intensity ratio may be attributed to the high sensitivity of the Raman modes to the laser polarization.

On the other hand, Zeng *et al.* investigated the Raman scattering of multilayer and monolayer MoS$_2$ in the low frequency regime (< 50 cm$^{-1}$).[22] For the wave length change of the laser from 532 to 633 nm, they observed the emergence of a broad Raman band around 38 cm$^{-1}$, which is pinned at a fixed energy, independent of the layer number, showing the complicated low-frequency Raman scattering of transition metal dichalcogenides. This phenomenon was interpreted as the consequence of the electronic Raman scattering associated with the spin-orbit coupling-based splitting in a conduction band at K points in the Brillouin zone.

In contrast to 2H-MX$_2$ family that are constructed on hexagonal lattices, naturally-formed WTe$_2$ has a distorted octahedral coordination about the tungsten atom, adopting an orthorhombic unit cell instead of a hexagonal one, and its symmetry is *Pmn21*.[23,24] The displacement of the tungsten atoms in this distorted phase increases the metallic bonding, and as a result tungsten chains are formed within the ditelluride planes along the *a*-axis of the orthorhombic unit cell, which makes this compound structurally one-dimensional and electronically a semimetal. Recently, it is shown that a single-crystalline WTe$_2$ possesses large, non-saturating magnetoresistance at cryogenic temperature[25], and it exhibits superconductivity under high pressure.[26,27] The discovery of extremely large positive magnetoresistance in WTe$_2$ presents new direction in the study of magnetoresistivity and can be used in potential applications of nanostructured memory devices and spintronics.[28] Such findings have stimulated enormous interest in other properties of WTe$_2$.

In this article, we present a systematic characterization on the Raman spectroscopic behavior of WTe$_2$ from bulk to monolayer using experimental and computational methods.[29] Special attention is placed on those of mono- to tri-layer WTe$_2$, which have not been explored until now, and we find that Raman spectroscopic analysis can offer a simple but robust method for their discrimination. We also find that most of the Raman modes of bulk WTe$_2$ consistently blueshift as the layer number decreases, while the frequency of the Raman mode associated with the lattice vibration along the tungsten chains remains almost invariant. Such Raman spectroscopic behaviors are totally different from those observed in hexagonal metal dichalcogenides. Density functional theory calculations validate these unique experimental results. Finally, we address the temporal degradation of Raman signals in WTe$_2$ films at ambient condition, suggesting a feasible Raman-based method to monitor the temporal corrosion of WTe$_2$.

## 2. Results and Discussion

### 2.1. Layer-Dependent Raman Spectra of WTe$_2$ Films

Atomically thin WTe$_2$ samples are isolated on SiO$_2$/Si substrates from commercial WTe$_2$ crystals by mechanical exfoliation method. **Figure 1**a shows the optical image of these WTe$_2$ samples deposited on the substrates. We determine the number of layers of WTe$_2$ samples by means of optical microscopy and atomic force microscopy (AFM) with a non-contact mode. Figure 1b is an AFM image obtained on the area marked by the red dashed rectangle in Figure 1a. The lateral thickness variation measured along the green solid line in Figure 1b indicates that a step height of one single layer in WTe$_2$ is 0.9–1.0 nm (Figure 1c), and thus the imaged flake in the scanned area is composed of mono- to tri-layer WTe$_2$ (see **Figure S1** in supporting information for optical and AFM images of other WTe$_2$ samples up to thirteen-layers). The step height of 0.9–1.0 nm in the AFM images is compatible with the interlayer spacing of ~0.75 nm in the WTe$_2$ crystals,[23] and the discrepancy between the AFM experiments and crystallography is due to the non-contact mode of AFM in the experiment.[30]

**Figure 2**a shows the evolution of the Raman spectra of WTe$_2$ as decreasing the thickness from bulk to monolayer. A typical Raman spectrum of bulk WTe$_2$ exhibits four distinct Raman peaks at ~118, ~134, ~164, and ~212 cm$^{-1}$ (hereafter they are denoted as $A_1^3$, $A_1^4$, $A_1^7$, and $A_1^9$, respectively, as defined in the next section) for the measurement above 100 cm$^{-1}$.

In contrast to 2H-MX$_2$ where the in-plane $E_{2g}^1$ mode stiffens and the out-of-plane $A_{1g}$ mode softens as the number of layers decreases,[14,15,18] we observe that all Raman modes of WTe$_2$ consistently blueshift except for the in-plane mode $A_1^7$, the frequency of which remains constant within ~1 cm$^{-1}$. For a bulk-to-monolayer transition, the most prominent frequency shift (~5 cm$^{-1}$) is achieved in the $A_1^9$ mode, while the $A_1^3$ and $A_1^4$ modes blueshift by ~4 and ~2 cm$^{-1}$, respectively (**Figure 2**a and **Figure 3**a).

Although a classical coupled harmonic oscillator model predicts phonon softening as the layer number decreases because interlayer interactions decrease effective restoring forces acting on the atoms,[31] WTe$_2$ should be quite free from this effect because this compound is rather structurally one-dimensional due to the formation of tungsten chains, which indicates that interlayer screening effects will dominate in determining the phonon frequencies of WTe$_2$. This fact explains the overwhelming blueshifts of the Raman modes of WTe$_2$ as the number of layers decreases. Meanwhile, the negligible shift of the $A_1^7$ mode is attributed to the fact that its lattice vibration occurs in the direction of tungsten chains (see the next section).

We also find that the Raman spectral patterns of mono- to tri-layer WTe$_2$ significantly differ from another. The four Raman peaks observed in bulk WTe$_2$ crystals are well-conserved in WTe$_2$ films down to trilayer. However, the Raman spectral pattern undergoes significant change in mono- and bi-layer WTe$_2$. That is, the Raman intensities of the $A_1^3$ and $A_1^4$ modes decays below the noise level in monolayer WTe$_2$ while the Raman intensity of the $A_1^4$ mode solely diminishes in bilayer WTe$_2$, exhibiting only two and three conspicuous Raman peaks, respectively (Figures 2a and 3b). These findings provide a simple and robust method to identify mono- and bi-layer WTe$_2$.

Very recently, the lattice dynamics of mono- and few-layer MoTe$_2$ were studied by Yamamoto *et. al.*[20] They found that the phonon modes of few-layer and sing-layer MoTe$_2$ have quite different optical activities to each other, and the Raman scattering of the $B_{2g}^1$ mode is enhanced in few-layer MoTe$_2$ while it is suppressed in monolayer MoTe$_2$. We suppose that the anomalous optical behaviors of the $A_1^3$ and $A_1^4$ modes of WTe$_2$ may have the same origin as the $B_{2g}^1$ mode of MoTe$_2$.

In order to get the clear picture of this phenomenon, we carry out the Raman intensity mapping of the WTe$_2$ sample shown in Figure 2b. Figures 2c and 2d show the Raman intensity mapping that is measured on the black-dotted rectangle in Figure 2b at the frequencies of ~134 and ~118 cm$^{-1}$ for the $A_1^4$ and $A_1^3$ modes, respectively. During this process, the Raman spectra have been measured within a total acquisition time of ~30 minutes because mono- and bi-layer WTe$_2$ are quickly aged within an hour, leading to the quenches of the Raman intensities as will be discussed in the last section. The Raman intensity map constructed for the $A_1^4$ mode indicates that no conceivable Raman scattering takes place in the mono- and bi-layer regions, but this scattering occurs intensively in the trilayer region (Figure 2c). Meanwhile, the Raman intensity map for the $A_1^3$ mode exhibits the strong Raman scattering in the bi- and tri-layer regions, but a conspicuous scattering is still lacking in the monolayer region (Figure 2d).

Another stimulating result is that the Raman intensities of the $A_1^7$ and $A_1^9$ modes are stronger in monolayer WTe$_2$ than in bulk crystals as shown in Figure 3b. The Raman intensities increase linearly up to trilayer WTe$_2$ and then decrease as the number of layers increases. Similar Raman phenomena were observed in graphene[32] and MoS$_2$[14] films deposited on SiO$_2$/Si substrates, where the maximum intensity occurs at approximately ten- and four-layers, respectively. As studied in graphene and MoS$_2$, we infer that this Raman intensity behavior results from the following three factors, the multiple reflection of Raman signals inside the layers, the interference effect arising from the multiple reflection of the excitation laser,[32] and the significant optical field enhancement caused by the interference effect between the oxide film and the WTe$_2$ layers.[14] The similar significant optical field enhancement was also observed in metal-MoS$_2$ heterostructures.[33]

Figure 3c shows the full width at half maximum (FWHM) of $A_1^7$ and $A_1^9$ modes plotted as a function of the number of layers. We find that the FWHMs increase from ~6 to ~7.6 cm$^{-1}$ for a mono- to bi-layer transition, and they monotonically decrease as the number of layers increases. This trend is the same as what observed in MoS$_2$[14], but the FWHM monotonically decreases with increasing the thickness in WS$_2$ and WSe$_2$.[19]

## 2.2. Computational Modelling of Raman Dynamics

For a deep understanding of these experimental observations described above, we have performed DFT calculations for the lattice vibrations and Raman scattering of WTe$_2$. The structure of bulk WTe$_2$ is shown in Figure 1d, which belongs to the space group *Pmn21*. It has 12 atoms in the orthorhombic unit cell, and there are 33 phonon modes (all of which are Raman active) at the Brillion zone center whose irreducible representation is expressed as follows:

$$\Gamma_{\text{bulk}} = 11A_1 + 6A_2 + 5B_1 + 11B_2 \qquad (1)$$

Meanwhile, monolayer WTe$_2$ belongs to the space group *P21/m* and contains 6 atoms in its period unit cell. Considering these, we also obtain the irreducible representation for the phonon modes (only 9 modes are Raman-active among 15 phonon modes) of monolayer WTe$_2$ as shown below:

$$\Gamma_{\text{mono}} = 6A_g + 2A_u + 3B_g + 4B_u \qquad (2)$$

The phonon spectra of bulk and monolayer WTe$_2$ are shown in **Figure 4**a and 4b, respectively. They exhibit similar phonon spectral patterns except for certain frequency shifts. The phonon spectra of bi- and tri-layer WTe$_2$ are shown in **Figure S2** (supporting information), respectively, exhibiting comparatively similar phonon spectral patterns as in bulk and monolayer WTe$_2$.

Finally, by calculating dielectric constant tensors over the atomic displacements, we obtain the Raman spectra of bulk and monolayer WTe$_2$. For the notation of Raman modes, we use $Z_m^n$ where $Z=A, B$; $m = 1, 2, g, u$; $n$ = positive integers listed in the order from the lowest frequency phonon mode. For bulk WTe$_2$, we obtain five distinct Raman peaks ($A_1^2$, $A_1^3$, $A_1^4$, $A_1^7$, and $A_1^9$). The lowest frequency mode ($A_1^2$) is not observed in our experiments since the Raman measurements have been performed above 100 cm$^{-1}$ (**Figure 5**a). The frequencies of these Raman modes are 74.65, 114.98 (118), 131.26 (134), 165.75 (164), and 209.48 (212) cm$^{-1}$, respectively, showing excellent agreement with the experimental results (given in parentheses). The lattice vibrations of these Raman modes are illustrated in Figure 5d. It indicates that the lattice vibration of the $A_1^7$ mode occurs in the direction of the tungsten chains.

In addition, it is worthy of noting that the Raman intensity extinctions of the $A_1^3$ and $A_1^4$ modes of monolayer WTe$_2$ are successfully captured in our DFT study (Figure 5b). However, we were not able to obtain the similar Raman-intensity ratio between the $A_1^7$ and $A_1^9$ modes as that in the experiment, presumably due to insufficient computational accuracy. This conjecture is supported by the fact that this discrepancy reduces when we use gradient-corrected approximation (GGA) instead of local-density approximation (LDA) in the calculations (Figure 5c). The DFT study also successfully reproduces the blueshifts of the $A_1^2$, $A_1^3$, $A_1^4$, and $A_1^9$ modes (by ~14.61, ~2.702, ~3.302, and ~8.473 cm$^{-1}$, respectively) as well as the negligible frequency change of the $A_1^7$ mode (within ~1.77 cm$^{-1}$) for a bulk-to-monolayer transition, showing good qualitative and quantitative agreements with the experimental observations. It is worth of noting that the $A_1^2$ mode that is predicted to emerge at ~75 cm$^{-1}$ tremendously blueshifts (by ~15 cm$^{-1}$) upon a bulk-to-monolayer transition, suggesting its valuable potential in the diagnosis of the number of layers in WTe$_2$.

**2.3. Temporal Degradation of Raman Signals**

**Figure 6**a shows the optical images obtained from the WTe$_2$ flake with an interval of 1 hour after it is freshly deposited onto the substrate. We find that the surface of the WTe$_2$ flake is severely corroded with time. Figure 6b shows Raman spectra acquired in the bilayer region of the WTe$_2$ flake at the same sequential stages as in Figure 6a. After one hour, the intensities of the Raman-active modes significantly degrade down to about one third of the initial values while the frequencies of the peaks remain almost constant. After two hours, their Raman intensities have completely vanished. This finding indicates that the temporal corrosion of WTe$_2$ can be effectively traced by monitoring the Raman signals. In contrast to bilayer WTe$_2$,

the Raman signals of trilayer WTe$_2$ remain fairly intact for several days (see **Figure S3** in supporting information for the optical images and Raman spectra of trilayer WTe$_2$).

Recently, black phosphorous, one of emerging layered materials, is shown to be very susceptible to atmospheric corrosion,[34-36] but it can be avoided using Al$_2$O$_3$ passivation layers.[37] Hence, in the same way, we infer that the surface of WTe$_2$ could be effectively protected against atmospheric corrosion using oxide passivation layers.

## 3. Conclusions

In summary, we have systematically investigated the layer-dependent lattice dynamics of WTe$_2$ from bulk to monolayer using optical microscopy, AFM, Raman spectroscopy, and DFT calculations. In contrast to hexagonal metal dichalcogenides in which both blue- and red-shifts of the Raman modes are observed, all Raman modes of WTe$_2$ consistently blueshift as decreasing the thickness down to monolayer except for the $A_1^7$ mode whose frequency remains almost constant within 1 cm$^{-1}$. Such anomalous Raman behaviors basically results from the distorted orthorhombic lattice of WTe$_2$. The structural distortion in WTe$_2$ leads to the formation of tungsten chains within the ditelluride planes, showing quasi-one-dimensional structural characteristics. The Raman spectrum of bulk WTe$_2$ exhibits four distinct peaks above 100 cm$^{-1}$, and their intensities are well-conserved down to trilayer. However, we discover that single peak has substantially decayed below the noise level in bilayer WTe$_2$, while double peaks vanish in monolayer WTe$_2$, suggesting a convenient and robust method to identify mono- and bi-layer WTe$_2$. The DFT calculations validate the above findings and predict the presence of low-frequency Raman mode at ~75 cm$^{-1}$ which drastically blueshifts (~15 cm$^{-1}$) for a bulk-to-monolayer transition. This study will contribute greatly to our understanding of the lattice dynamics of atomically-thin WTe$_2$, pointing to significant impacts of the distorted phase on the lattice dynamics of layered materials in two-dimensional regime.

## 4. Experimental Section

### 4.1. Preparation of Mono- and Few-layer WTe$_2$

A WTe$_2$ flake with mono and few-layer regions are exfoliated from commercial bulk WTe$_2$ (2D semiconductors Inc.) using micro-mechanical method, and then deposited onto a silicon substrate covered by a 300 nm thick SiO$_2$ layer. The size of monolayer WTe$_2$ ranges typically from ~1 to 3 μm. This size is much smaller than that of typical graphene and other TMDs. As can be seen in Figure 1a, however, monolayer regions of a WTe$_2$ flake are easily characterized by high optical visibilities in the optical microscopy images. The numbers of layers of the WTe$_2$ flake are further determined by non-contacted AFM.

### 4.2. Raman Spectroscopy Measurements

Micro Raman spectroscopy technique is employed to investigate the layer-dependent Raman dynamics of WTe$_2$. The lights are collected and analyzed in the back scattering geometry in ambient conditions using a 532 nm solid-state laser, 100x objective with 1.25 N.A. (numerical aperture), and a grating with 1800 grooves per millimeter. The laser power is kept below 40 μW during the measurements to avoid possible heating and sample damages.[38] The instrumental spectral resolution is ~1 cm$^{-1}$, and the first order Raman scattering in Si (~520 cm$^{-1}$) is used as a frequency reference.

### 4.3. Density Functional Theory Calculations

The DFT calculations are performed within LDA (GGA is used in the case of the accurate dielectric matrix calculation of monolayer WTe$_2$) using projector-augmented wave potentials and Perdew–Burke–Ernzerhof exchange–correlation parameterization as implemented in the Vienna *ab initio* simulation package (VASP).[39] The kinetic energy cutoff of the plane-wave basis is set to be 340 eV and 24×14×6 Monkhorst Pack grid is used for k-space sampling in dielectric matrix calculations. The calculated Hellmann–Feynman forces and dielectric matrix are imported into PHONON [40] to generate their derivatives for the construction and diagonalization of dynamic and Raman matrices.

**Supporting Information**

**Acknowledgements**

This work was supported by the KIST Institutional Program (Project No. 2E25382) and also by the Center for Advanced Meta-Materials (CAMM) funded by the "Ministry of Science, ICT and Future Planning" as Global Frontier Project (CAMM-No. 2014M3A6B3063727). We would like to thank Sohyun Park in Advanced Analysis Center at KIST for her support in Raman experiments.

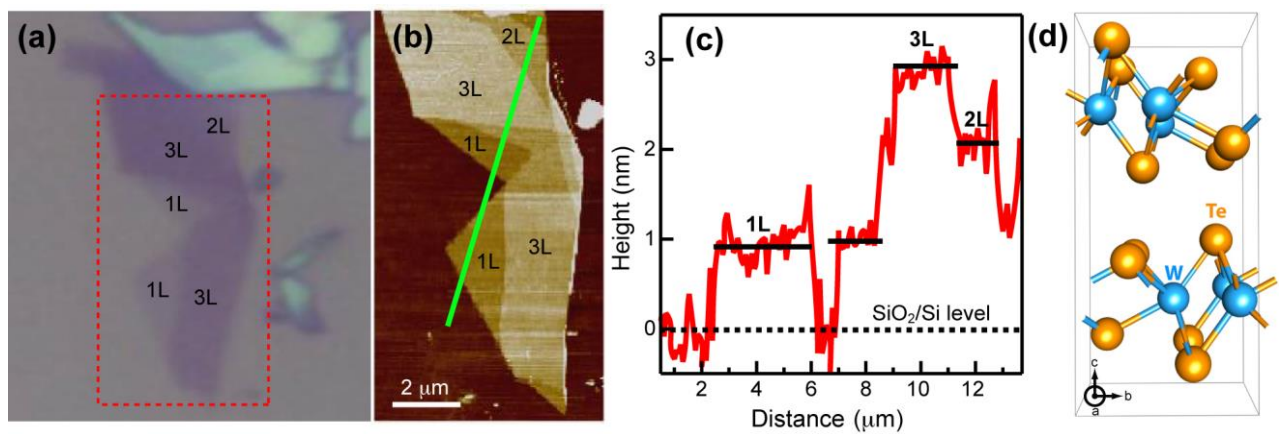

**Figure 1.** (a) A typical optical image of a WTe$_2$ flake (containing mono- to tri-layer regions) deposited on Si/SiO$_2$ substrate. The number of layers is indicated by $n$L with $n$ = 1–3; (b) AFM image of the area marked by the red dashed rectangle in (a); (c) A height profile of a WTe$_2$ flake measured along the green solid line in (b), indicating a step height of 0.9–1 nm for a single layer of WTe$_2$; (d) The structure of WTe$_2$ crystals.

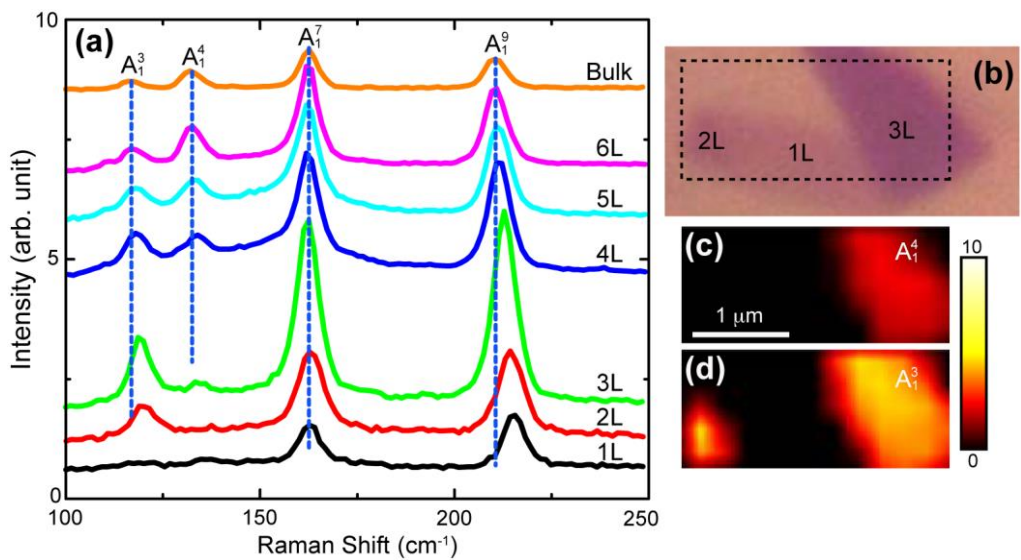

**Figure 2.** (a) The Raman spectra of bulk and mono- to six-layer WTe$_2$. The blue dashed lines indicate the frequencies of the Raman modes in bulk WTe$_2$; (b) An optical image of a WTe$_2$ flake containing mono- to tri-layer regions; (c,d) The Raman intensity maps for the $A_1^4$ (top) and $A_1^3$ (bottom) modes measured in the black dotted region in (b).

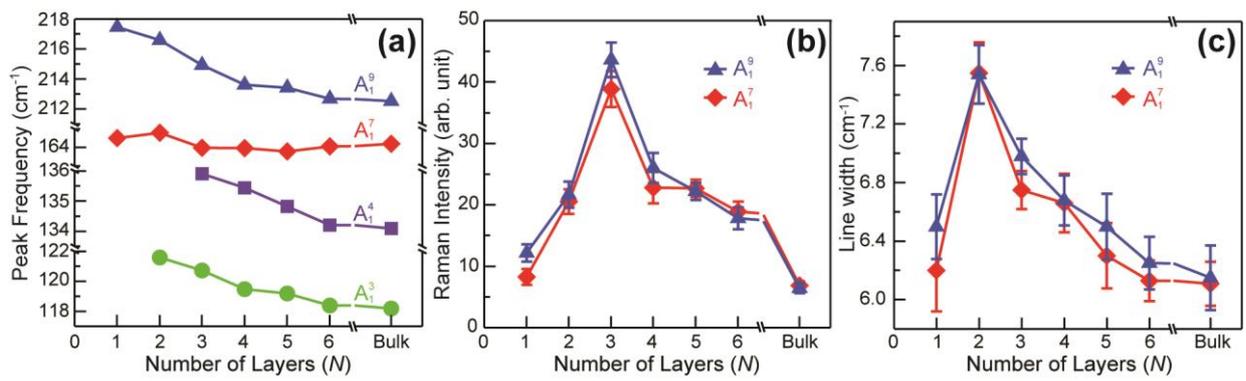

**Figure 3.** (a) The Raman peak frequencies of WTe$_2$ plotted as a function of the number of layers; (b) The intensity and (c) line-width variations of the $A_1^7$ and $A_1^9$ modes plotted as a function of the number of layers.

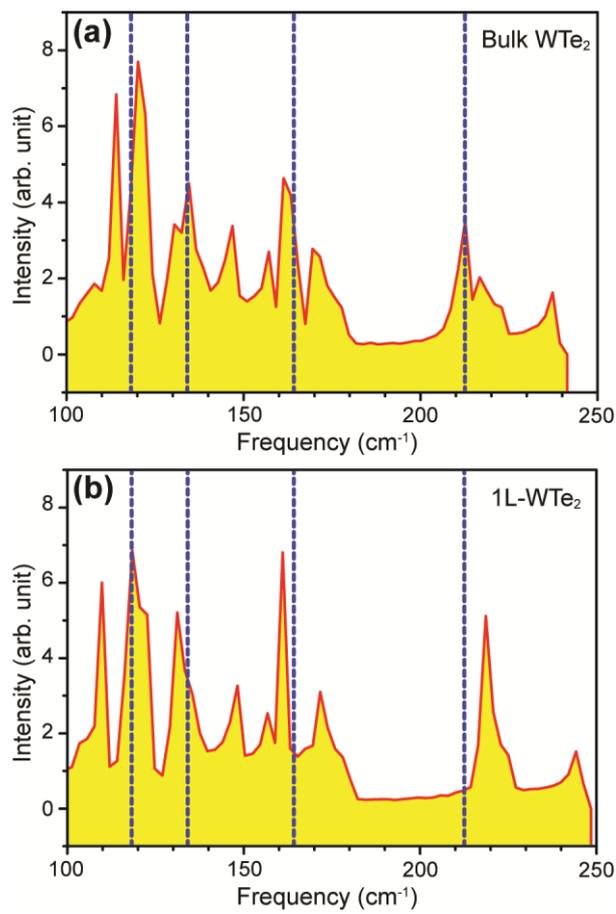

**Figure 4.** The phonon spectra of (a) bulk and (b) monolayer $WTe_2$. Blue dashed lines indicate the lattice vibrational frequencies of the Raman modes in bulk $WTe_2$ measured in the experiment.

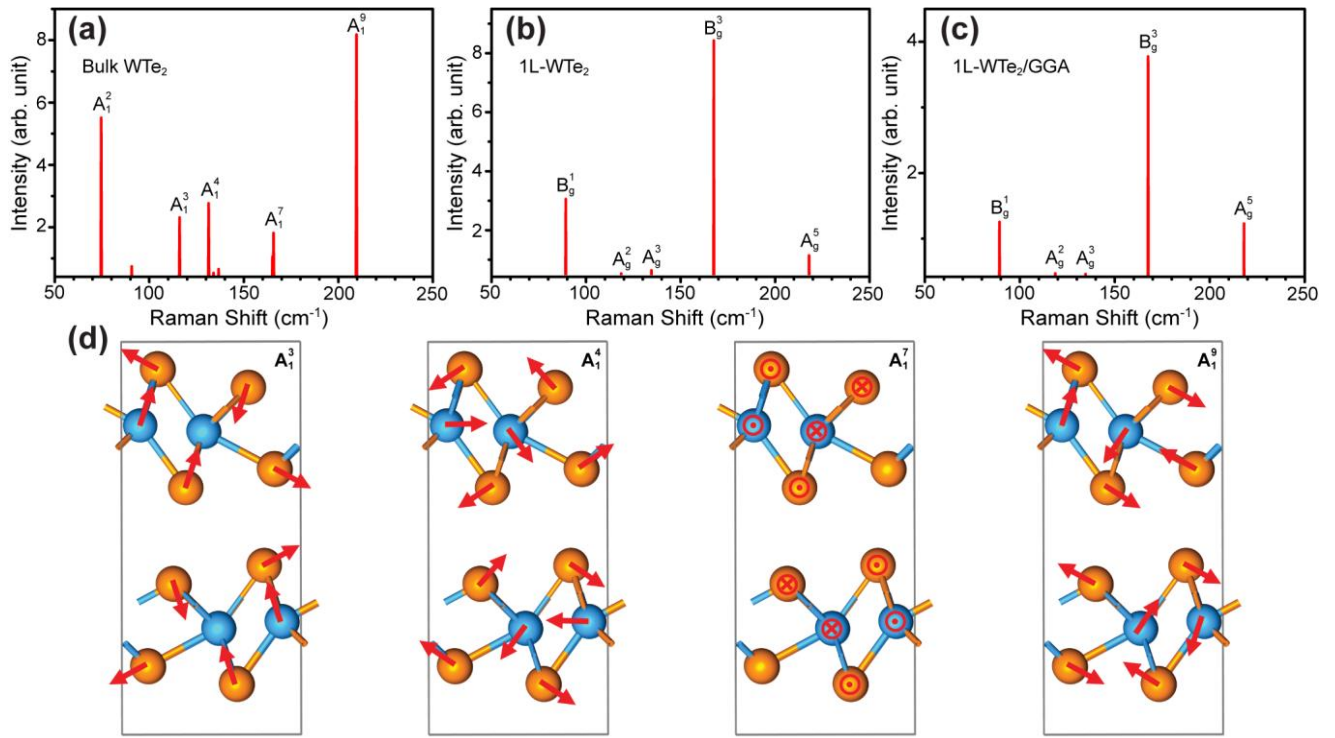

**Figure 5.** The Raman spectra of (a) bulk and (b) monolayer WTe$_2$ obtained from LDA-based DFT calculations, and (c) the Raman spectrum of monolayer WTe$_2$ obtained from GGA-based DFT calculation. The bottoms are offset by 10% of the maximum intensity to consider noise cancellation; (d) Atomic illustration for the lattice vibrations of the Raman modes in bulk WTe$_2$.

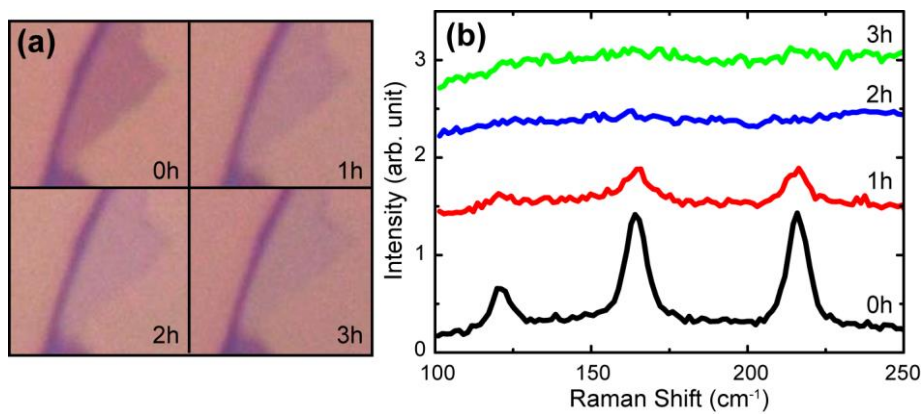

**Figure 6.** (a) The optical images of bilayer $WTe_2$ acquired at sequential stages after its fresh deposition on $Si/SiO_2$ substrate. (b) The Raman spectra of bilayer $WTe_2$ measured on the center of a degraded region at the same sequential stages as in (a).

**Info for table of contents**

The evolution of the Raman spectrum of WTe$_2$ from bulk to monolayer (left). The optical image of a WTe$_2$ flake containing mono- to tri-layer regions and the Raman intensity maps of the $A_1^3$ and $A_1^4$ modes (middle). Atomic visualization for the lattice vibration of the $A_1^7$ mode which is shown to occur in the direction of tungsten chains (right).

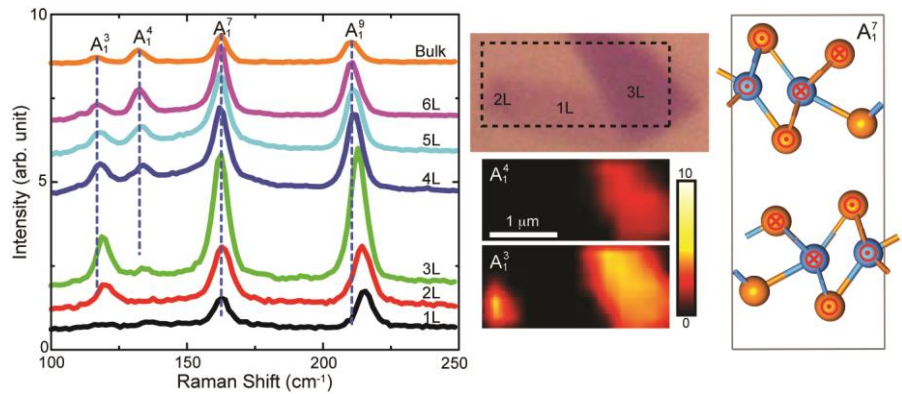

Supporting Information for

# *Anomalous Lattice Dynamics of Mono-, Bi-, and Tri-layer WTe$_2$*

Younghee Kim,[†] Young In Jhon,[†] June Park,[†] Jae Hun Kim, Seok Lee, and Young Min Jhon*

Sensor System Research Center, Korea Institute of Science and Technology, Seoul 136-791, Republic of Korea

[†]These authors contributed equally to this work.

# Optical microscopy and AFM images of 5- to 13-layer WTe$_2$

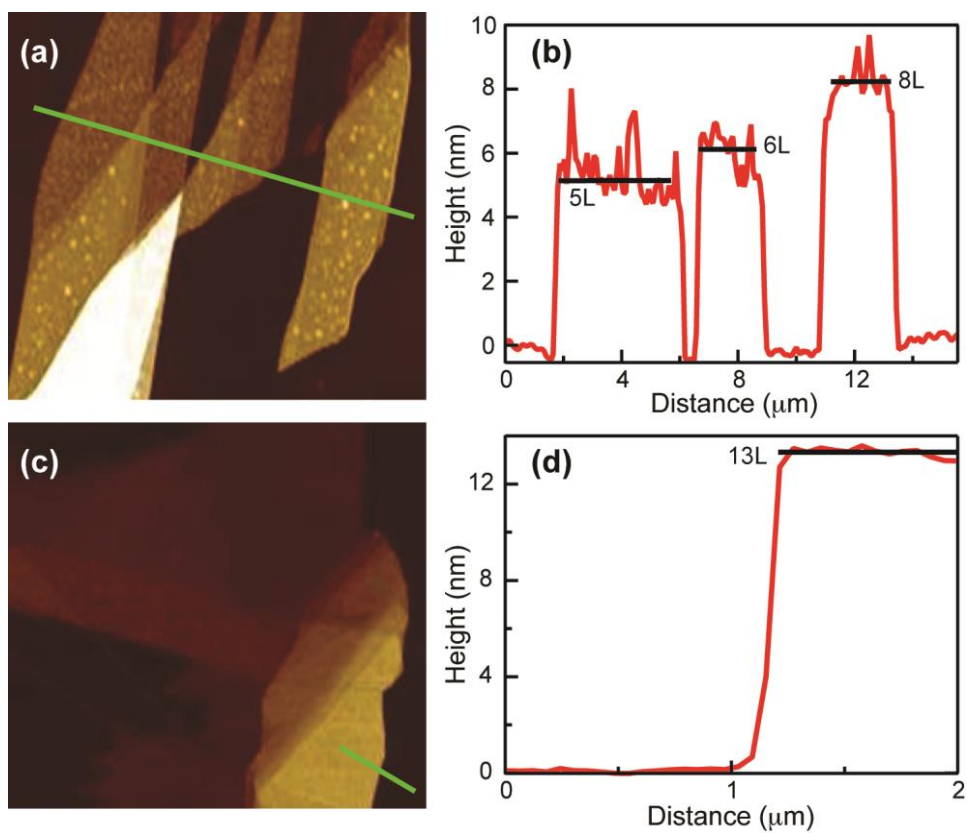

**Figure S1.** (a, c) The optical microscopy images of WTe$_2$ flakes containing "5, 6, and 8 layers" and "13 layers", respectively; (b, d) the AFM images of the WTe$_2$ flakes measured along the green solid lines in (a, c).

# The phonon spectra of bulk and mono- to tri-layer WTe$_2$

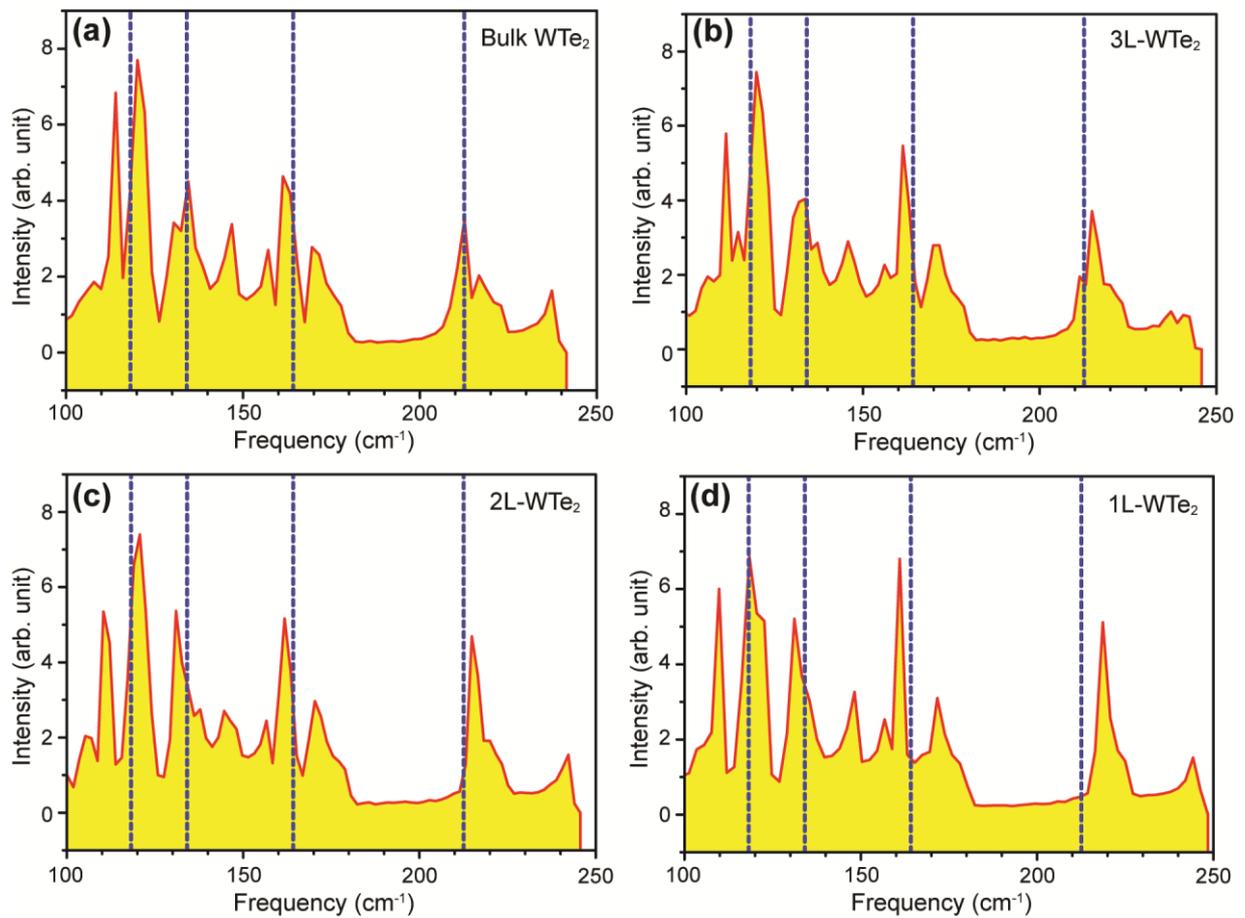

**Figure S2.** Simulated Phonon spectra of (a) bulk, (b) tri-layer, (c), bi-layer, and (d) mono-layer WTe$_2$. Blue dashed lines indicate lattice vibrational frequencies corresponding to Raman modes of bulk WTe$_2$ in experiments.

# Temporal degradation of trilayer WTe$_2$

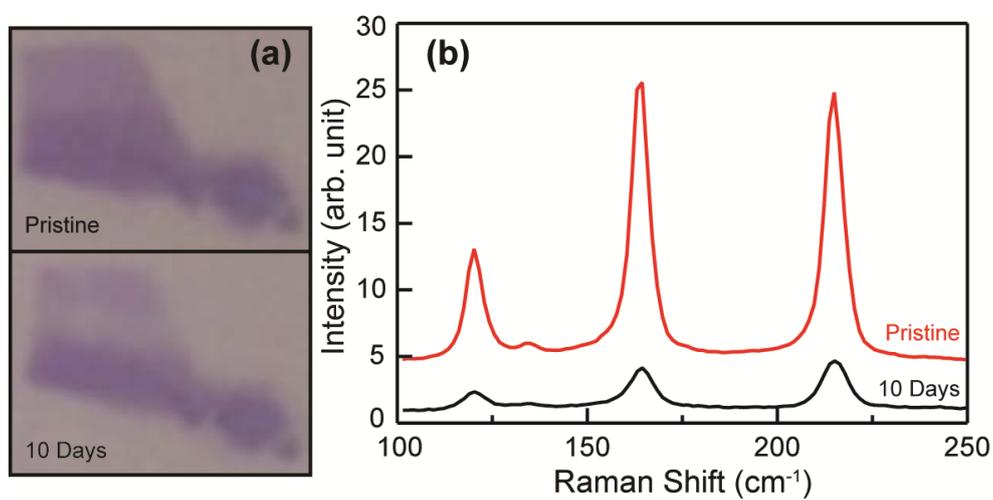

**Figure S3.** (a) The optical microscopy images of trilayer WTe$_2$ obtained at sequential stages after its fresh deposition on Si/SiO$_2$ substrate. (b) The Raman spectra of trilayer WTe$_2$ measured on the center of degraded regions at the same sequential stages as in (a).